\documentstyle[eqsecnum,multicol,aps]{revtex}
\begin{document}
\preprint {}
\title {Nonequilibrium Phase Transition in the Kinetic Ising model:\\
 Dynamical symmetry breaking by randomly varying magnetic field}
\author {Muktish Acharyya$^*$}
\address {Institute for Theoretical Physics,\\
University of Cologne, 50923 Cologne, Germany}
\date{\today}
\maketitle
\begin{abstract}
The nonequilibrium dynamic phase transition,
in the two dimensional kinetic Ising model 
in presence of a randomly varying (in time but uniform in space)
 magnetic field,
has been studied both by Monte Carlo
simulation and by solving the mean field dynamic equation of motion for
the average magnetisation. 
In both the cases,
the time averaged magnetisation vanishes
from a nonzero value depending
upon the values of the width of randomly varying field and the temperature.
The phase boundary lines are drawn in the plane formed by the width of the
random field and the temperature.

\noindent{PACS number(s): 05.50.+q}
\end{abstract}

\begin{multicols}{2}

\section{Introduction}
The problem of the random field Ising model (RFIM), having {\it quenched} 
random field, has been investigated
both theoretically \cite{imry,aha,fisher,young} and experimentally
\cite{king} in the last few years 
because it helps to simulate many interesting but complicated
problems. The effects of a randomly {\it quenched} magnetic field 
on the critical behaviour near the ferromagnetic phase transition is the
special focus of the modern research. The recent developments in this field can
be found in a review article \cite{springer}.
However, the dynamical aspects of the Ising system in presence of a randomly
varying field is not yet studied
thoroughly. It would be interesting to know if there is any dynamical phase
transition in the presence of a randomly varying magnetic field.

For completeness and continuity,
it would be convenient to review briefly the previous studies on the dynamic
transition in the kinetic Ising model.
Tome and Oliviera \cite{tom} first 
observed and studied the dynamic transition in the
kinetic Ising model in presence of a sinusoidally oscillating magnetic field. 
They solved the mean field (MF)
dynamic equation of motion (for the average magnetisation) of the kinetic
Ising model in presence of a sinusoidally oscillating magnetic field.
By defining the order parameter as the time averaged
magnetisation over a full cycle of the 
oscillating magnetic field they showed that 
the order parameter vanishes
depending upon the value of the temperature and the 
amplitude of the oscillating field. 
Precisely, in the field amplitude and temperature plane they have drawn
a phase boundary separating dynamic ordered 
(nonzero value of the order parameter) and disordered (order 
parameter vanishes) phases. They \cite{tom} have also observed 
and located a {\it tricritical point} (TCP),
[separating the nature (discontinuous/continuous) of the transition]
on the phase boundary line. 
However, such a transition, observed \cite{tom} from the solution of mean field
dynamical equation, is not dynamic in true sense.
This is because, for the field amplitude less than the
coercive field (at temperature less than the transition temperature
without any field), the response magnetisation varies periodically
but asymmetrically even in the zero frequency limit; the system remains locked
to one well of the free energy and cannot go to the other one, in the absence
of noise or fluctuation.

Lo and Pelcovits \cite{lo} first attempted to study the dynamic 
nature of this phase transition (incorporating the effect of fluctuation)
in the kinetic Ising model by Monte Carlo (MC) simulation. 
In this case, the transition disappears in the 
zero frequency limit; due to the 
presence of fluctuations, the magnetisation flips to the
direction of the magnetic field and the dynamic order parameter (time
averaged magnetisation) vanishes.
However, they \cite{lo} have 
not reported any precise phase boundary.
Acharyya and Chakrabarti \cite{ac} studied the nonequilibrium dynamic phase
transition in the kinetic Ising model 
in presence of oscillating magnetic field by 
extensive MC simulation. They \cite{ac} have successfully drawn the phase
boundary for the dynamic transition and observed/located a tricritical 
point on it.
It was also noticed by them \cite{ac}
that this dynamic phase transition is associated with the 
breaking of the symmetry of
the dynamic hysteresis ($m-h$) loop. 
In the dynamically disordered (value of order parameter vanishes)
phase the corresponding hysteresis loop is
symmetric, and loses its symmetry in the ordered phase (giving
nonzero value of dynamic order parameter).
They 
have \cite{ac} also studied the temperature variation of the AC susceptibility
components near the dynamic transition point.  
They observed 
that the imaginary (real) part of the AC susceptibility gives a
peak (dip) near the dynamic transition point (where the dynamic
order parameter vanishes). They \cite{ac} have the following 
conclusions: (i) this is a distinct
signal of a phase transition and (ii) this is an indication of the
thermodynamic nature of the phase transition.

Recently, 
the relaxation behaviour, of the dynamic order parameter,
near the transition point, has been studied
\cite{ma1} both
by MC simulation and solving meanfield dynamic equation. It has been observed 
that the relaxation is Debye type and the relaxation time diverges near the
transition point, showing a {\it critical slowing down}.
 The 'specific heat' and the 'susceptibility' also diverge
\cite{ma2}
near the transition point in a manner similar to that of fluctuations of
the dynamic order parameter and energy respectively.

The statistical distribution of the dynamic
order parameter, near the dynamic transition point,
has been studied by Sides et al \cite{rik}. They observed that
the distribution widens up (to a double hump from a single hump type)
as one crosses the dynamic transition phase boundary.
Since, the fluctuation increases as the 
width of a distribution increases,
this observation is consistent with the independent observation \cite{ma2}
of critical fluctuations of dynamic order parameter.
They \cite{rik} have
also observed 
that the fluctuation of the hysteresis loop area becomes considerably
large near the dynamic transition point.

In this paper, the dynamic phase transition has been studied, in the 
two dimensional 
kinetic Ising model in the 
presence of randomly varying (in time but uniform in
space) magnetic field, both
by MC simulation and by solving the meanfield dynamical equation. 
This paper
is organised as follows: in section II the model and the MC simulation scheme
with the results are given, in section III the meanfield dynamical equation
and its solution with the numerical results are given and the paper ends
with a summary of the work in section IV.

\section {Monte Carlo study}
\subsection{The Model and the simulation scheme}

The Hamiltonian, of an Ising model (with ferromagnetic nearest
neighbour interaction) in presence of a time varying magnetic field, can
be written as
\begin{equation}
H = -\sum_{<ij>} J_{ij} s_i^z s_j^z - h(t) \sum_i s_i^z.
\label{hm} 
\end{equation}
Here, $s_i^z (=\pm 1)$ is Ising spin variable, $J_{ij}$ is the interaction
strength and
$h(t)$  
is the randomly varying (in time but uniform in space) magnetic field. The
time variation of $h(t)$ can be expressed as

\begin{equation}
h(t) = \left\{
\begin{array}{rcl}
h_0 r(t) &  ~~{\rm for}~~ & t_0 < t < t_0 +\tau \\
      0  & ~~{\rm elsewhere}~~ &
\end{array}\right.
\label{field}
\end{equation}
\noindent where $r(t)$ is a random variable distributed uniformly
between -1/2 and +1/2. The field $h(t)$ varies randomly from $-h_0/2$ to
$h_0/2$ and

\begin{equation}
{1 \over {\tau}} \int_{t_0}^{t_0+\tau} h(t) dt = 0.
\end{equation}

The system
is in contact with an isothermal heat bath at temperature $T$. For simplicity
the values of all $J_{ij}$ are taken equal to 
unity. The periodic boundary condition is used here.

A square lattice of linear size $L (=100)$ has been considered.
Initially all spins are taken directed upward and $h(t) = 0$.
At any finite
temperature $T$,
 the 
dynamics of this system has been studied here by Monte Carlo
simulation using Metropolis single spin-flip dynamics. The transition rate 
 ($s_i^z \to -s_i^z$) is specified as
\begin{equation}
W(s_i^z \to -s_i^z) = {\rm Min}\left[1, \exp(-{\Delta H}/K_BT)\right]
\end{equation}
\noindent where $\Delta H$ is the change in energy due to spin flip and
$K_B$ is the Boltzmann constant which has been taken unity here for
simplicity.
Each lattice site is
updated here sequentially and one such full scan over the lattice is
defined as the time unit (Monte Carlo step per spin or MCSS)
here. The magnitude of the field $h(t)$ changes after every MCSS obeying
equation \ref{field}. 
The instantaneous magnetisation
(per site),
 $m(t) = (1/L^2) \sum_i s_i^z $ has been calculated. 
After bringing the system in steady state ($m(t)$ gets stablised with some
fluctuations), the switch
of the randomly varying magnetic field $h(t)$ has been turned on 
(at time $t_0$ MCSS) and calculated
the instanteneous magnetisation. The $t_0$ has been taken equal
to $2\times10^6$ and even more 
($3.15\times10^6$)
near the static ferro-para transition
temperature ($\sim 2.269..$). By inspecting the data and the time variation, 
it is observed that $m(t)$ gets stabilised, for this choice of the value of
$t_0$.
The time averaged (over the active period of the magnetic field)
magnetisation $Q = {1 \over {\tau}}
 \int_{t_0}^{t_0+\tau} m(t) dt$ has been calculated over sufficiently
large time $\tau$ ($1.75\times 10^6$). By changing the values of $\tau$ 
the stabilisation of $Q$ has been checked quite
carefully. 
It is observed that,
$Q$ does not change much (apart from the small fluctuations)
 and this choice of the value of
$\tau$ is considered to be good enough to believe that the value of $Q$
becomes stable. 
Here, $Q$ plays the role of dynamic order parameter.
It may be noted that,
one measures the value of order parameter i.e., sponteneous magnetisation,
 in the same
way in the case of normal ferro-para transition. But, there is a remarkable
difference. In that case, system reaches a steady equilibrium state, however
in this case the state lies in a {\it nonequilibrium} (time dependent) state.
This dynamic order parameter $Q$ is observed to be a function of width
($h_0$) of the randomly
varying magnetic field and the temperature ($T$) of the system, i.e., 
$Q = Q(h_0, T)$. Each value of $Q$ has been calculated by averaging over 
at least 25 
different random samples.

\subsection{Results}

Taking all spins are up ($s_i^z = +1$) as the initial condition, the
simulation was performed using the above form of the time varying field
$h(t)$. 
It has been observed numerically that,
for a fixed values of $h_0$, if $T$ increases, $Q$
decreases continuously 
and ultimately vanishes at a fixed value of $T$. Similarly,
for a fixed temperature, if $h_0$ increases, the value of $Q$ 
decreases and finally vanishes at a particular value of $h_0$. Figure 1
shows the time variation of magnetisation $m(t)$ at a particular temperature
$T$ and for two different values of field width $h_0$. For a small value
of $h_0$, the system remains in a dynamically symmetry broken phase
(Fig. 1a) where
the magnetisation oscillates asymmetrically about zero line. As a result,
the dynamical order parameter $Q$ (the time averaged magnetisation) is nonzero.
As the field increases, the system gets
sufficient energy to flip dynamically in such a way that the magnetisation
oscillates symmetrically (Fig. 1b) about zero line and as a result $Q$
vanishes.

It is possible to let $Q$ vanish, either by increasing the temperature $T$
for a fixed field width $h_0$ or vice versa. 
It has been observed that for any fixed value of $h_0$ the transition is
continuous. Two such transitions (for two different values of $h_0$) is
depicted in Fig. 2.
So, in the plane formed by
the temperature ($T$) and the field width ($h_0$), one can think of a
boundary line, below which $Q$ is nonzero and above
which it vanishes. Figure 3 displays such a phase boundary in the $h_0-T$
plane obtained numerically by Monte Carlo simulation. The transition observed
here is continuous irrespective of the values of $h_0$ and $T$, i.e., 
unlike the earlier cases \cite{ac,tom}, no
{\it tricritical point} has been observed. It may be noted here, 
that the transition
temperature, for field width going to zero, reduces to the equilibrium 
(zero field) ferro-para transition point. In fig. 3, the overestimated 
(from the Onsager value $T_c = 2.269..$) value of the 
transition temperature in $h_0 \to 0$ limit, 
may be due to the small size of the system. It should
be mentioned here, that all
the results are obtained by using sequential
updating scheme. 
Although, there is some study \cite{rik1}
regarding the 
updating scheme to simulate the dynamic processes, 
in the present study no significant deviation
was found for these two different (sequential/random) updating techniques.
Few data points of the phase boundary (in Fig. 3 marked by bullets)
are obtained by
using the random updating scheme and no significant deviation was observed.

\section {Meanfield study}

\subsection {Meanfield dynamical equation and solution}

The meanfield dynamical equation of motion for the average magnetisation
($m$) \cite{tom} is
\begin{equation}
{dm \over dt} = -m + {\rm tanh}\left({{m(t) + h(t)} \over T}\right)
\label{mf}
\end{equation}
\noindent where $h(t)$ is the 
randomly varying external magnetic field satisfying the following condition
\begin{eqnarray}
{1 \over {\tau}} \int_{t_0}^{t_0+\tau} h(t) dt = 0,
\end{eqnarray}
\noindent with the same distribution $P(h)$ discussed earlier.

The equation \ref{mf} has been solved emplyoing the fourth order Runge-Kutta
method subjected to the above condition. The initial
magnetisation is set equal to unity, which serves as a boundary condition.
First,
the magnetisation $m(t)$ has been calculated for $h(t)$ = 0 and brought the
system into an equilibrium state and after that the switch of the 
randomly varying magnetic field
has been turned on. The dynamic order parameter $Q = {1 \over \tau }
\int_0^{\tau}m(t') dt'$
has been calculated, from the solution $m(t)$ of equation \ref{mf}.
The $Q$ is averaged over 25 different random samples.
\bigskip
\subsection {Results}

Observations similar to the MC case 
are made in this case. For quite small values
of $h_0$ and $T$ systems remains in a dynamically asymmetric phase 
($Q \neq 0$) and gets into
a dynamically symmetric 
($Q = 0$) phase for higher values of $h_0$ and $T$.
It is observed that one can force $Q$ to vanish continuously 
by tuning $h_0$ and $T$. A phase 
boundary line for the dynamic transition is shown in Fig. 4.
Here also, the limiting ($h_0 \to 0$) trantistion temperature reduces to the
MF equilibrium (zero field) transition point ($T_c^{MF} = 1$). It may be noted
here that the coordination number $z$ (= 4 in two dimension) has been 
absorbed in the interaction strength $J$ in calculating
the $T_c^{MF} ( = 1)$ and the temperature (shown here in Fig. 4) is measured
in the unit of $Jz$.

\bigskip
\section{Summary}

The nonequlibrium dynamic phase transition, in the two dimensional
kinetic Ising model
in presence of a randomly varying
(in time but uniform over the space) magnetic field, is studied both by Monte
Carlo simulation and by solving the meanfield dynamical equation of motion.
In both the cases, it is observed that the
system remains in a dynamically symmetric phase 
($Q = 0$) for large values of the width
of the randomly varying magnetic field and the temperature. By reducing the
value of field width and temperature one can bring the system in a dynamically
symmetry broken phase ($Q \neq 0$). The 
 time averaged magnetisation, i.e.,
the dynamic order parameter vanishes continuously  depending upon
the value of width of the randomly varying field and the temperature. The 
nature of the transition observed here is 
always continuous and the phase boundaries are drawn. It may be mentioned
here that the dynamic responses, of Glauber kinetic Ising model, are studied 
\cite{ac} for a quasiperiodic time variation
of the magnetic field.

There is some experimental evidence of dynamic transition. Recently, Jiang
et al \cite{jiang}
observed the 
indications of dynamic transition, 
associated with the dynamical symmetry breaking,
in the ultrathin Co/Cu(001) sample put
in a sinusoidally oscillating magnetic field by
magneto-optic Kerr effect.
For small values of the amplitude of the oscillating field the $m-h$ loop
lies asymmetrically (dynamic ordered phase)
in the upper half plane and becomes symmetric 
(dynamic disordered phase) for higher values of the
field amplitude. Very recently the dynamical symmetry breaking has been observed
experimentally in highly anisotropic (Ising like) ultrathin ferromagnetic
Fe/W(110) films and nicely depicted in Fig. 1 of Ref. \cite{suen}.
However, the detailed quantitative study
to draw the phase boundary is not yet done experimentally.

Very recently, the stationary properties 
of the Ising ferromagnet in presence of
a randomly varying (having bimodal distribution) magnetic
field are studied \cite{rujan}
in the meanfield approximation. The transition observed
from the distribution of stationary magnetization
is discontinuous. 
However this present study deals with the
{\it dynamical} properties (dynamical symmetry breaking) 
of the system and the
nonequilibrium transitions in the Ising ferromagnet in presence of randomly
varying (uniformly distributed) magnetic field.
In the former case the transition was observed from the distribution of 
stationary magnetisation. However, in the present study, this was observed
from the temperature variation of dynamic order parameter associated with
a dynamical symmetry breaking and the transition observed is continuous.
An extensive numerical effort is required to characterise and to know
the details of this
dynamical phase transition. It would also be important to see whether the
different kind of 
distribution of the randomly varying field would give different results.

\section*{Acknowledgements}

Sonderforschungsbereich 341 is gratefully acknowledged for financial support.
The author would like to thank D. Stauffer and B. K. Chakrabarti 
for careful reading of the
manuscript.

\vskip 0.3 cm

\centerline {\bf Figure captions}

\bigskip

\noindent {\bf Fig. 1.} 
Monte Carlo results of the
time ($t$) variations of randomly varying field $h(t)$
and the response magnetisation $m(t)$ for $T$ = 1.7 and (a) $h_0$ = 1.0
and (b) $h_0$ = 3.0. The symmetry (about the zero line) breaking is clear
from the figure.

\bigskip

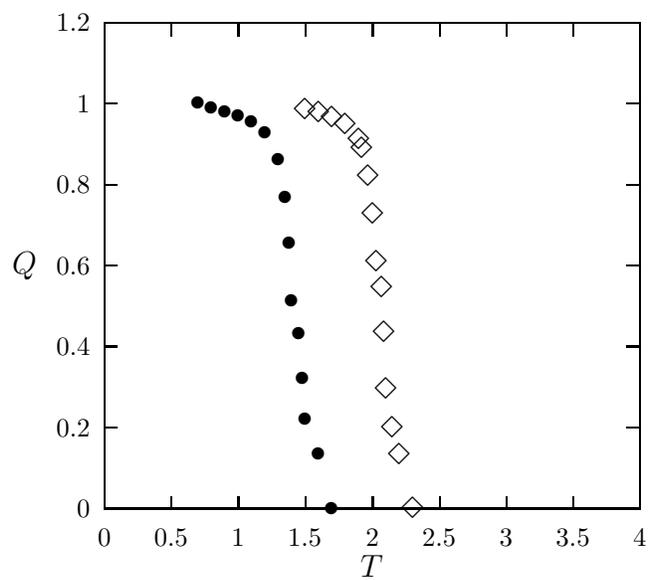
\noindent {\bf Fig. 2.} Monte Carlo results of the temperature 
($T$) variations
of the dynamic order parameter $Q$ for two different values of $h_0$.
($\bullet$) for $h_0$ = 2.4 and ($\Diamond$) for $h_0$ = 0.8.

\bigskip

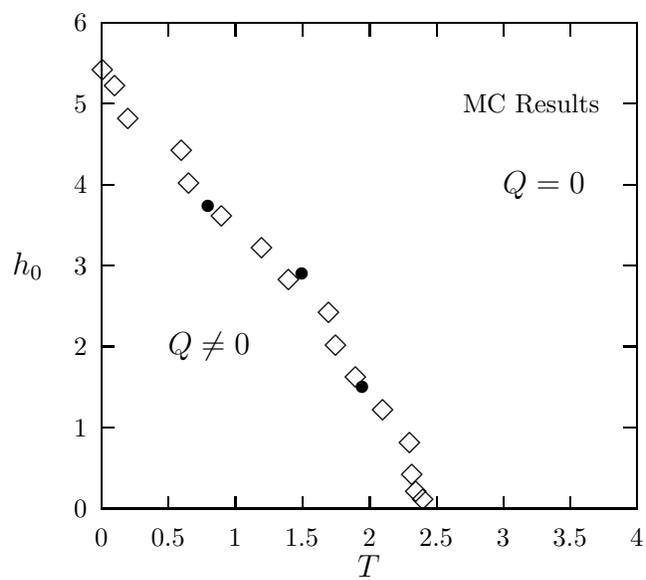
\noindent {\bf Fig. 3.} The phase boundary for the dynamic transition obtained
from MC simulations. The data points represented by ($\bullet$) are obtained
by using random updating and those represented by ($\Diamond$) are obtained
by sequential updating.

\bigskip

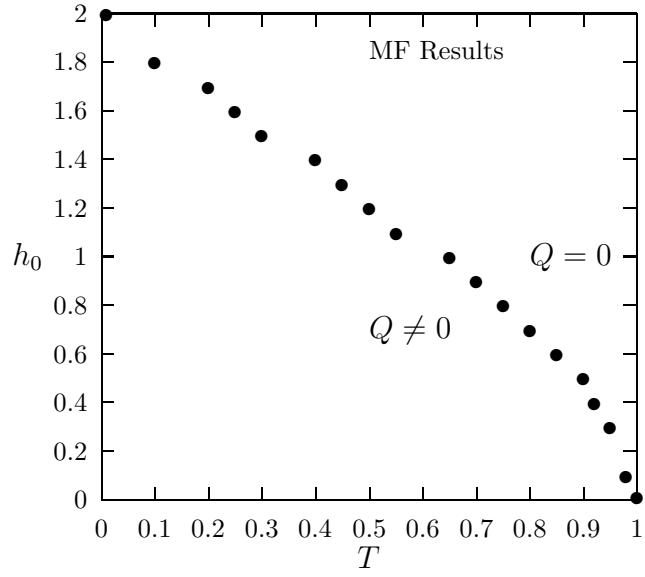
\noindent {\bf Fig. 4.} The phase boundary for the dynamic transition obtained
from the solution of the MF dynamic equation (equation \ref{mf}). 
\end{multicols}
\end{document}